\documentclass[twocolumn,tighten]{aastex61}
\pdfoutput=1 
\usepackage{amsmath,amstext}
\usepackage[T1]{fontenc}
\usepackage{apjfonts} 
\usepackage[figure,figure*]{hypcap}

\shorttitle{Anti-correlated X-ray and Radio Variability in PSR J1023+0038}
\shortauthors{Bogdanov et al.}
\submitjournal{The Astrophysical Journal}

\begin{document}

\title{Simultaneous Chandra and VLA Observations of the Transitional Millisecond Pulsar PSR~J1023+0038: Anti-correlated X-ray and Radio Variability}

\author{Slavko Bogdanov}
\affiliation{Columbia Astrophysics Laboratory, Columbia University, 550 West 120th Street, New York, NY 10027, USA}
\author{Adam T.~Deller}
\affiliation{Centre for Astrophysics and Supercomputing, Swinburne University of Technology, Hawthorn, VIC 3122, Australia}
\author{James C.~A.~Miller-Jones}
\affiliation{International Centre for Radio Astronomy Research - Curtin University, GPO Box U1987, Perth, WA 6845, Australia}
\author{Anne M.~Archibald}
\affiliation{Anton Pannekoek Institute for Astronomy, University of Amsterdam, Science Park 904, NL-1098 XH Amsterdam, The Netherlands}
\affiliation{ASTRON, Netherlands Institute for Radio Astronomy, Postbus 2, NL-7990 AA Dwingeloo, The Netherlands}
\author{Jason W.~T.~Hessels}
\affiliation{Anton Pannekoek Institute for Astronomy, University of Amsterdam, Science Park 904, NL-1098 XH Amsterdam, The Netherlands}
\affiliation{ASTRON, Netherlands Institute for Radio Astronomy, Postbus 2, NL-7990 AA Dwingeloo, The Netherlands}
\author{Amruta Jaodand}
\affiliation{Anton Pannekoek Institute for Astronomy, University of Amsterdam, Science Park 904, NL-1098 XH Amsterdam, The Netherlands}
\affiliation{ASTRON, Netherlands Institute for Radio Astronomy, Postbus 2, NL-7990 AA Dwingeloo, The Netherlands}
\author{Alessandro Patruno}
\affiliation{Leiden Observatory, Leiden University, P.O. Box 9513, NL-2300 RA Leiden, The Netherlands}
\author{Cees Bassa}
\affiliation{ASTRON, Netherlands Institute for Radio Astronomy, Postbus 2, NL-7990 AA Dwingeloo, The Netherlands}
\author{Caroline D'Angelo}
\affiliation{Leiden Observatory, Leiden University, P.O. Box 9513, NL-2300 RA Leiden, The Netherlands}

\begin{abstract}
We present coordinated \textit{Chandra X-ray Observatory} and Karl G.~Jansky Very Large Array observations of the transitional millisecond pulsar PSR~J1023+0038 in its low-luminosity accreting state. The unprecedented five hours of strictly simultaneous X-ray and radio continuum coverage for the first time unambiguously show a highly reproducible, anti-correlated variability pattern. The characteristic switches from the X-ray high mode into a low mode are always accompanied by a radio brightening with duration that closely matches the X-ray low mode interval. This behavior cannot be explained by a canonical inflow/outflow accretion model where the radiated emission and the jet luminosity are powered by, and positively correlated with, the available accretion energy.  We interpret this phenomenology as alternating episodes of low-level accretion onto the neutron star during the X-ray high mode that are interrupted by rapid ejections of plasma by the active rotation-powered pulsar, possibly initiated by a reconfiguration of the pulsar magnetosphere, that cause a transition to a less luminous X-ray mode. The observed anti-correlation between radio and X-ray luminosity has an additional consequence: transitional MSPs can make excursions into a region of the radio/X-ray luminosity plane previously thought to be occupied solely by black hole X-ray binary sources. This complicates the use of this luminosity relation to identify candidate black holes, suggesting the need for additional discriminants when attempting to establish the true nature of the accretor. 
\end{abstract}

\section{Introduction}

PSR~J1023$+$0038 is the archetypal ``missing link" binary millisecond pulsar \citep{2009Sci...324.1411A}.  The binary system contains a $P=1.69$ ms pulsar and a $0.2-0.5$ M$_{\odot}$ secondary star \citep{2012ApJ...756L..25D,2015MNRAS.451.3468M} in a 4.8 hour orbit. It is the first system to show evidence for alternating radio-loud pulsar and X-ray binary states \citep{2005AJ....130..759T,2009ApJ...703.2017W}. The system underwent a transformation to a low-luminosity accreting state in 2013 June \citep{2014ApJ...781L...3P,2014ApJ...790...39S}, in which it has remained until present.

In the low-luminosity accreting state, PSR J1023$+$0038 exhibits puzzling X-ray modulation characterized by rapid switching between two clearly distinguishable luminosity levels (referred to as ``modes'') of $\approx$$3\times10^{33}$ and $\approx$$5\times10^{32}$ ergs s$^{-1}$ (0.3--10 keV), plus sporadic flares \citep{2015ApJ...806..148B,2014ApJ...791...77T}. This unusual large-amplitude variability appears to be a generic feature of accreting transitional millisecond pulsars \citep[tMSPs, see][]{2013A&A...550A..89D,2014MNRAS.438..251L,2015ApJ...803L..27B}.
 \textit{XMM-Newton} observations also revealed
coherent X-ray pulsations at the 1.7-ms pulsar rotation period known from the radio pulsar state \citep{2015ApJ...807...62A}, which strongly suggest active accretion at very low luminosities ($\sim$$10^{33}$ ergs s$^{-1}$). 
This discovery comes as a surprise since the implied
accretion rate ($\sim$$10^{-5}-10^{-4}$ times the Eddington luminosity) is
expected to be too low to overcome the ``centrifugal barrier'' imposed
by the pulsar's $\sim$$10^{8}$ G magnetic field and rapid spin.
Unlike luminous accreting millisecond X-ray pulsars \citep[AMXPs; see][for a comprehensive overview]{2012arXiv1206.2727P}, the pulsations are only observed in
the steady ``high'' mode (in
which the system spends $\sim$75\% of its time) and they seem to
switch off during the ``low'' mode (20\% of the time) and in the bright
flares ($1-2$\% of the time).  Furthermore, the X-ray high mode flux and
pulse shape are so stable that they are nearly identical in 
observations spaced years apart \citep{2015ApJ...806..148B,2016ApJ...830..122J}.  The presence of channeled
accretion at a very specific accretion rate and the peculiar X-ray flux 
bimodality are very puzzling and cannot be satisfactorily explained by
existing accretion models.

A long-term X-ray timing study with \textit{XMM-Newton} \citep{2016ApJ...830..122J} revealed that, in its accreting state, PSR J1023$+$0038 is spinning down $26.8\pm0.4$\% faster compared to its previous disk-free radio pulsar state. The implication of this finding is that accretion does not dramatically alter the pulsar spin-down rate in this state, and that the pulsar wind is still active at a similar (albeit potentially not identical) level as in the disk-free rotation-powered state. 

At radio frequencies, observations have shown that in its accreting state PSR J1023$+$0038 is not detectable as a radio pulsar \citep{2014ApJ...790...39S}. However,  VLA radio imaging during this state revealed rapidly variable (with factor-of-two variability within  minutes and order-of-magnitude
variability on timescales of $\sim$30 minutes) flat-spectrum emission that persisted over a period of at least 6 months \citep{2015ApJ...809...13D}. 
Using near-simultaneous radio and X-ray observations, \citet{2015ApJ...809...13D} found that this emission is consistent with a compact, partially self-absorbed synchrotron jet, and that 
PSR J1023$+$0038 (and other tMSPs) exhibit much stronger jets than  predictions based on observations of neutron stars accreting at much higher rates.

At present, it is not clear if/how the radio variability is causally
related to the X-ray behavior. In particular, the jet production
mechanism might operate in only one or two of these X-ray modes;
alternatively, a jet may be present in all modes but with substantially changing spectral and temporal properties.  Previous coordinated X-ray and radio observations only resulted in a brief interval of simultaneous coverage, which proved inconclusive in establishing any correlated behavior between the two wavelength bands \citep{2015ApJ...806..148B}.

In this paper, we present an analysis of strictly simultaneous X-ray and radio continuum observations of PSR J1023+0038 during the current accreting state. The results provide important new information regarding the puzzling X-ray and radio behavior, and their relation. The work is organized as follows. In Section 2 we discuss the X-ray and radio observations and the data reduction procedures. In Section 3 we present a variability and correlation analysis of the X-ray and radio time series. In Section 4, we focus on the X-ray/radio luminosity relation in light of the new results for PSR J1023+0038. We discuss the implications of the results in Section 5 and offer concluding remarks in Section 6.

\begin{figure*}[t!]
\centering
\includegraphics[angle=0,width=0.75\textwidth]{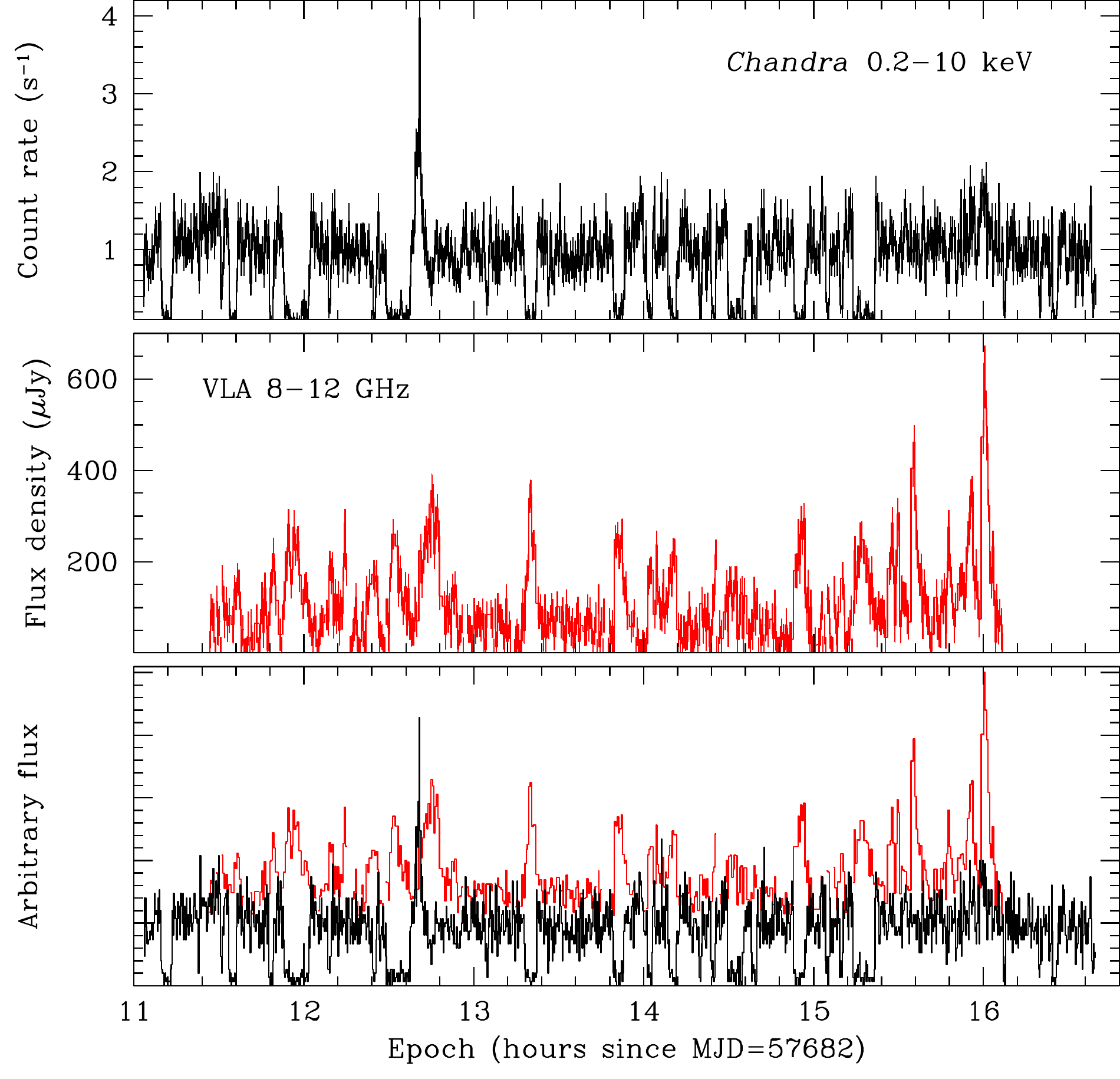}
\caption{\label{fig:full}{\bf Top:} \textit{Chandra} 0.2--10 keV (black) exposure-corrected and background subtracted count rate time series of PSR J1023+0038 grouped in 25 s intervals. The data shows the three distinct X-ray flux modes (low, high, and flare) seen in previous observations. {\bf Middle:} VLA 8--12 GHz light curve binned at a 30 s resolution.    {\bf Bottom:} The same light curves from the upper panels superposed, with the error bars omitted to improve clarity. The increased radio activity associated with each X-ray low mode is obvious. }
\end{figure*}

\begin{figure}[t!]
\includegraphics[angle=0,width=0.42\textwidth]{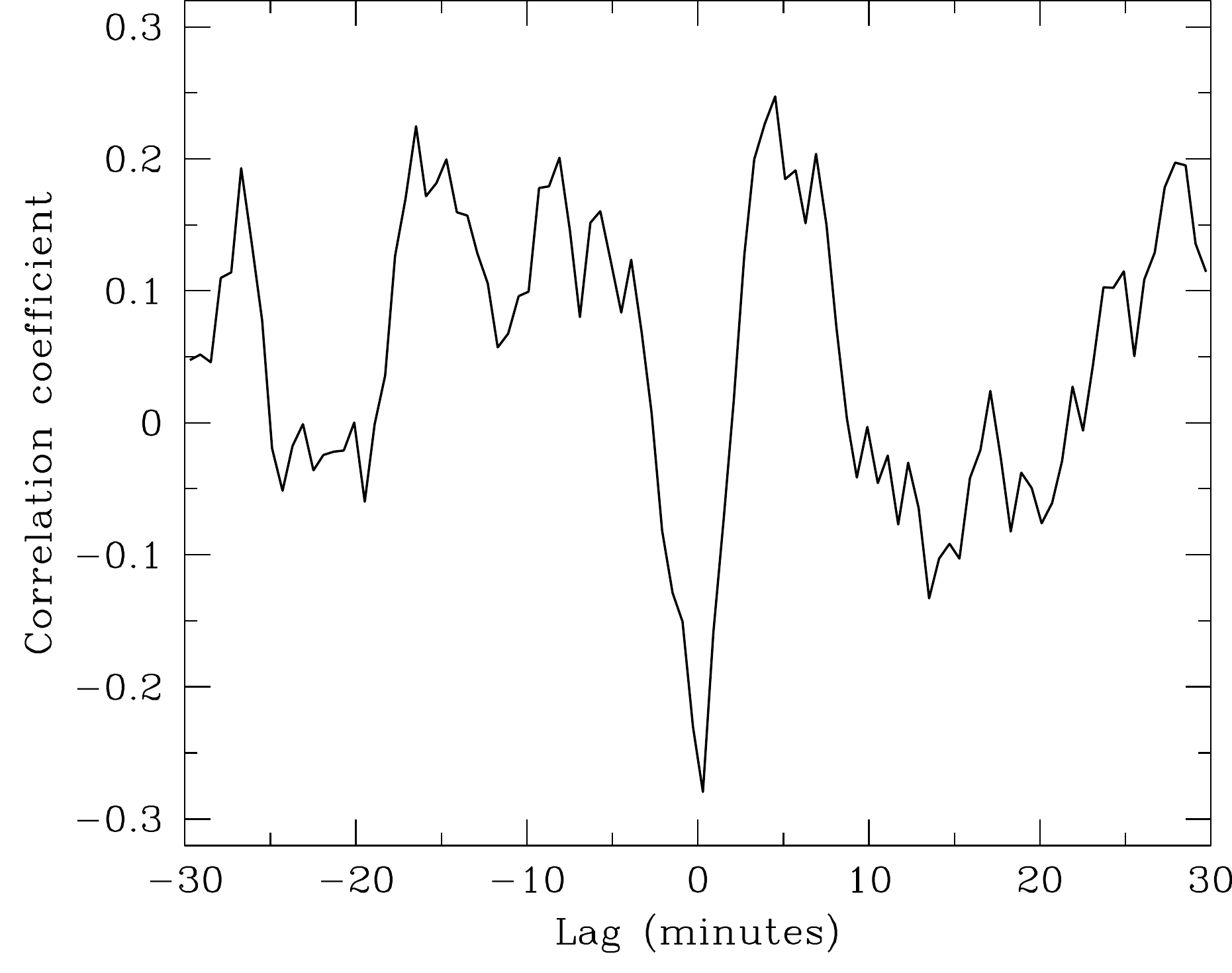}
\caption{\label{fig:xcf} Cross-correlation of the \textit{Chandra} and VLA time series. A positive lag corresponds to the X-ray leading the radio in time. A strong anti-correlation near zero lag is evident.}
\end{figure}

\section{Observations and Data Analysis}
\subsection{Chandra X-ray Observatory}
PSR J1023+0038 was observed with \textit{Chandra} as part of the Cycle 17 Guest Observer program (ObsID 17785) on 2016 October 21 for a duration of 20 ks. The back-illuminated ACIS-S3 chip was placed at the aim point. To avoid photon pile-up\footnote{Event pile-up in CCDs occurs when two or more events fall on a detector pixel during the same readout frame. As a consequence, the intrinsic source spectrum is distorted and degraded. This deleterious effect can be mitigated by increasing the readout rate.} that would be caused by the high source count rate, the detector was used in Continuous Clocking (CC33\_GRADED) mode, which enables a rapid (2.85 ms) readout time by sacrificing one dimension of spatial imaging.

The data analysis was carried out using CIAO\footnote{Chandra Interactive Analysis of Observations.} version 4.9 \citep{2006SPIE.6270E..1VF} and the accompanying calibration database CALDB 4.7.3. Following recommended procedures, the source and background events were extracted from rectangular regions of width 5$''$ along the imaging direction. Due to the bright nature of PSR J1023+0038, the background contribution to the source count rate is negligible.

The event time stamps were translated from the Terrestrial Time (TT) standard used by \textit{Chandra} to UTC in order to enable alignment with the radio light curve.  As the read out time of the \textit{Chandra} data is longer than the pulsar spin period, a pulsation/timing analysis is not possible.

\subsection{Very Large Array}
A five-hour observation with the Karl G.~Jansky Very Large Array (VLA) was performed on 2016 October 21 from UT 11:16 to 16:15, entirely overlapping with the \textit{Chandra} observation. The array was in the A configuration and the observational setup and calibration was identical to that described in  \citet{2015ApJ...809...13D}, sampling a 4 GHz bandwidth from 8--12 GHz, using 3C286 as the primary flux density calibrator, and using J1024--0052 as the gain calibrator with a cycle time of 8 minutes. The VLA pipeline (in CASA version 4.5.0) was used for initial calibration, followed by a small amount of additional flagging on the target field and self-calibration using the field model described in \citet{2015ApJ...809...13D}.  The self-calibration procedure included the peeling (subtraction from the visibilities after direction-dependent calibration) of the brightest background source in the field.

Imaging was performed in an automated fashion in small windows that surround PSR J1023+0038 and several other background sources in the field, with selection criteria applied to produce a time series binned at 30 and 60 s time resolution over the entire observation, using the full frequency bandwidth.

In addition, we produced a number of stacked data sets that give higher sensitivity and hence allow a more detailed analysis of flux density and spectral index variations. For these analyses, we produced single-frequency and dual-frequency (8--10 and 10--12 GHz) datasets for plotting, and also datasets with five equally-spaced frequency segments to better estimate the spectral index and its error.  First, we stacked all the time ranges corresponding to 1) clear X-ray high modes, and 2) clear X-ray low modes.  We stacked the $\sim$4 minute intervals that correspond to the rising and decaying phase of the radio flare, which follows the sole bright X-ray flare at UT 12:40.  Finally, we stacked time ranges corresponding to X-ray low modes of similar duration, using the high$\rightarrow$low mode transition as a reference point for alignment.  This allowed us to compare long (duration $\sim$7.7 minutes), medium (duration $\sim$3.3 minutes), and short (duration $\sim$1.5 minutes) low modes.

The source flux density was extracted using the CASA task \textit{imfit} with a starting model fixed at the known source position, while the image noise was estimated using an emission-free region of the image offset by several arcseconds from the source.  Failed fits and non-detections were replaced by upper limits at twice the image noise.

\subsection{Archival Data}
In Section 3, we revisit the contemporaneous \textit{XMM-Newton} and VLA observations from 2013 November 10 originally presented in \citet{2015ApJ...806..148B} and \citet{2015ApJ...809...13D} (see those papers for the details of the observations and data reduction). In previous analyses of the \textit{XMM-Newton} data, we only considered the European Photon Imaging Camera (EPIC) MOS1/2 and PN instruments, which overlapped for 16 minutes with the VLA observation. However, due to a shorter instrumental setup time, the Reflection Grating Spectrometer (RGS) exposure begins 30 minutes earlier, providing a longer interval of simultaneity with the radio time series.

We used the  \textit{XMM-Newton} Science Analysis System\footnote{The \textit{XMM-Newton} SAS is developed and maintained by the Science Operations Centre at the European Space Astronomy Centre and the Survey Science Centre at the University of Leicester.} (SAS) version {\tt xmmsas\_20170112\_1337-16.0.0} to process the RGS data. A combined RGS1 and RGS2, first and second order, exposure-corrected and background-subtracted light curve binned at 20 seconds was produced using the {\tt rgslccorr} tool in SAS.

\begin{figure}[t!]
\includegraphics[angle=0,width=0.46\textwidth]{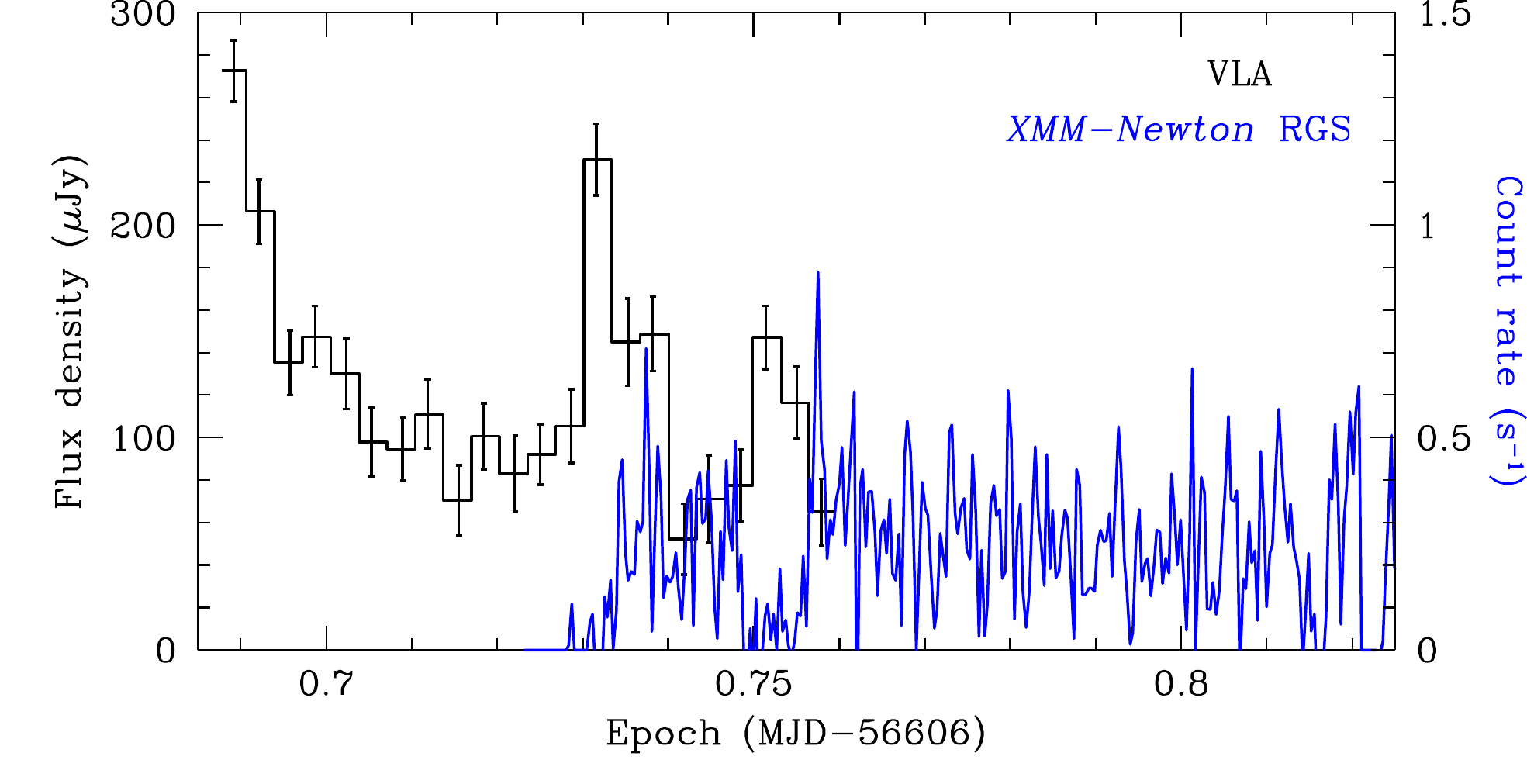}
\caption{\label{fig:rgs} Contemporaneous VLA 5.9 GHz (black) and \textit{XMM-Newton} RGS (blue) time series of PSR J1023+0038 obtained on 2013 November 10. The anti-correlated variability is evident, with a significant increase in radio flux during the two X-ray low modes that occur during the overlap.}
\end{figure}

\section{Variability Analysis}
The joint \textit{Chandra}/VLA observations resulted in $\sim$5 hours of strictly simultaneous coverage. The acquired exposure-corrected X-ray light curve in the $0.2-10$ keV range binned in 25 s intervals and the radio light curve in the $8-12$ GHz band with a 30 s time resolution are shown in Figure~\ref{fig:full}. The bin sizes were chosen to provide an optimal balance between signal-to-noise ratio (S/N) and time resolution. The X-ray light curve exhibits the characteristic variability pattern seen in previous observations \citep{2014ApJ...781L...3P,2014ApJ...790...39S,2015ApJ...806..148B,2016ApJ...830..122J}, namely, rapid and aperiodic switching between two distinct flux levels at $\approx$$5\times10^{32}$ (the low mode) and $\approx$$3\times10^{33}$ ergs s$^{-1}$ (the high mode) plus a short-lived flaring episode at UT 12:40 reaching $\sim$$10^{34}$ ergs s$^{-1}$.

The VLA time series shows numerous rapid flares that generally last a few minutes, interspersed between periods of relatively steady faint emission.  A striking anti-correlated variability pattern between the X-ray and radio is immediately apparent. Nearly every X-ray low mode interval is accompanied by a temporally coincident radio brightening, while the steady low-level radio emission occurs during the X-ray high modes. The ``quiescent'' radio flux during the X-ray high modes has an average flux density at the reference frequency of 10 GHz of $S_{\nu}=56 \pm 2$ $\mu$Jy and a spectral index of $\alpha= 0.2 \pm 0.2$ (where $S_{\nu}\propto \nu^{\alpha}$). A delayed radio flare follows the brief but intense X-ray flare around UT 12:40, with an average radio flux density at 10 GHz of $235\pm9$ $\mu$Jy.

Apart from the radio brightening seen during X-ray low modes, and following the one well-defined X-ray flare, several other radio flares that are not clearly associated with any particular X-ray activity are seen, mostly during the window UT $\sim$15:30 to 16:00.  The peak radio luminosity over the entire observation is actually attained during one of these ``unassociated'' radio flares.  Some weak X-ray flaring appears to be present around UT 15:50 to 16:00 (see Figure~\ref{fig:full}), but several of the radio flares with no X-ray counterparts occur prior to this window of weak X-ray flaring activity.

As expected based on visual inspection of the light curves, a cross-correlation analysis indicates a strong anti-correlation near zero lag between the X-ray and radio time series (Figure~\ref{fig:xcf}).  The archival \textit{XMM-Newton} RGS and VLA data from 2013 November 10 show the same anti-correlation for the two low-mode intervals covered in the $\sim$45 minute segment with simultaneous data (see Figure~\ref{fig:rgs}),  though due to the brief overlap, the results were previously inconclusive when considering this data set in isolation.

One of the remarkable features of the variability is that the rise and fall of the radio luminosity seems to very closely match the transitions to and from the X-ray low mode. Specifically, the radio brightening associated with the shorter duration low modes has correspondingly faster rise and decay times. Furthermore, there does not appear to be a strong dependence of the peak radio luminosity on the duration of a low mode; some of the short and medium duration low mode intervals reach a comparable radio luminosity to the longest low modes (see Figure~\ref{fig:long_flares}).

\subsection{Time-resolved Spectral Variations}
As found in previous studies \citep{2015ApJ...806..148B,2016A&A...594A..31C}, the X-ray spectrum of PSR J1023$+$0038 only undergoes a modest change between the three modes. The spectrum is power-law-like with spectral photon index $\Gamma\approx1.7$ at all luminosity levels.  

Looking at the average radio spectral index for PSR J1023$+$0038, as shown in Table~\ref{tab:radiospectrum}, suggests similarly minor variations between X-ray modes: $\alpha= 0.2 \pm 0.2$ in the high mode, $\alpha= -0.1 \pm 0.2$ averaged across all low modes, and $\alpha= 0.3 \pm 0.4$ during the radio flare following the prominent X-ray flare at UT 12:40.  However, averaging masks considerable variation {\em during} each radio brightening associated with an X-ray low mode.   While the sensitivity of the VLA data is insufficient to resolve each individual low mode both spectrally and temporally, stacking allows us to investigate the average radio spectral evolution during a low mode.  As shown in Table~\ref{tab:radiospectrum}, the radio spectrum evolves from significantly inverted ($\alpha \simeq 0.4$) to relatively steep ($\alpha \simeq -0.5$) over the course of several minutes during a typical low mode.

\begin{deluxetable}{lcc} 
\tablecolumns{2} 
\tablecaption{\label{tab:radiospectrum} Radio flux density and spectral index of PSR J1023$+$0038}
\tablehead{ 
\colhead{Selection criterion}    &  
\colhead{$S_{\mathrm{10GHz}}$ ($\mu$Jy)} & 
\colhead{Spectral index $\alpha$}  
}
\startdata 
X-ray high mode (all) 		& $56 \pm 2$	& $0.2 \pm 0.2$ \\
X-ray low mode (all)		& $168 \pm 4$	& $-0.1 \pm 0.2$ \\
X-ray low mode (1st half)	& $196 \pm 6$	& $0.4 \pm 0.3$ \\
X-ray low mode (2nd half)	& $156 \pm 6$	& $-0.5 \pm 0.3$ \\
Radio flare F1 (all) 		& $235 \pm 9$	& $0.3 \pm 0.4$ \\
Radio flare F1 (1st half) 	& $224 \pm 13$	& $0.4 \pm 0.5$ \\
Radio flare F1 (2nd half) 	& $260 \pm 11$	& $0.1 \pm 0.5$ \\
\enddata
\end{deluxetable}

\begin{figure*}[t!]
\centering
\includegraphics[angle=0,width=0.75\textwidth]{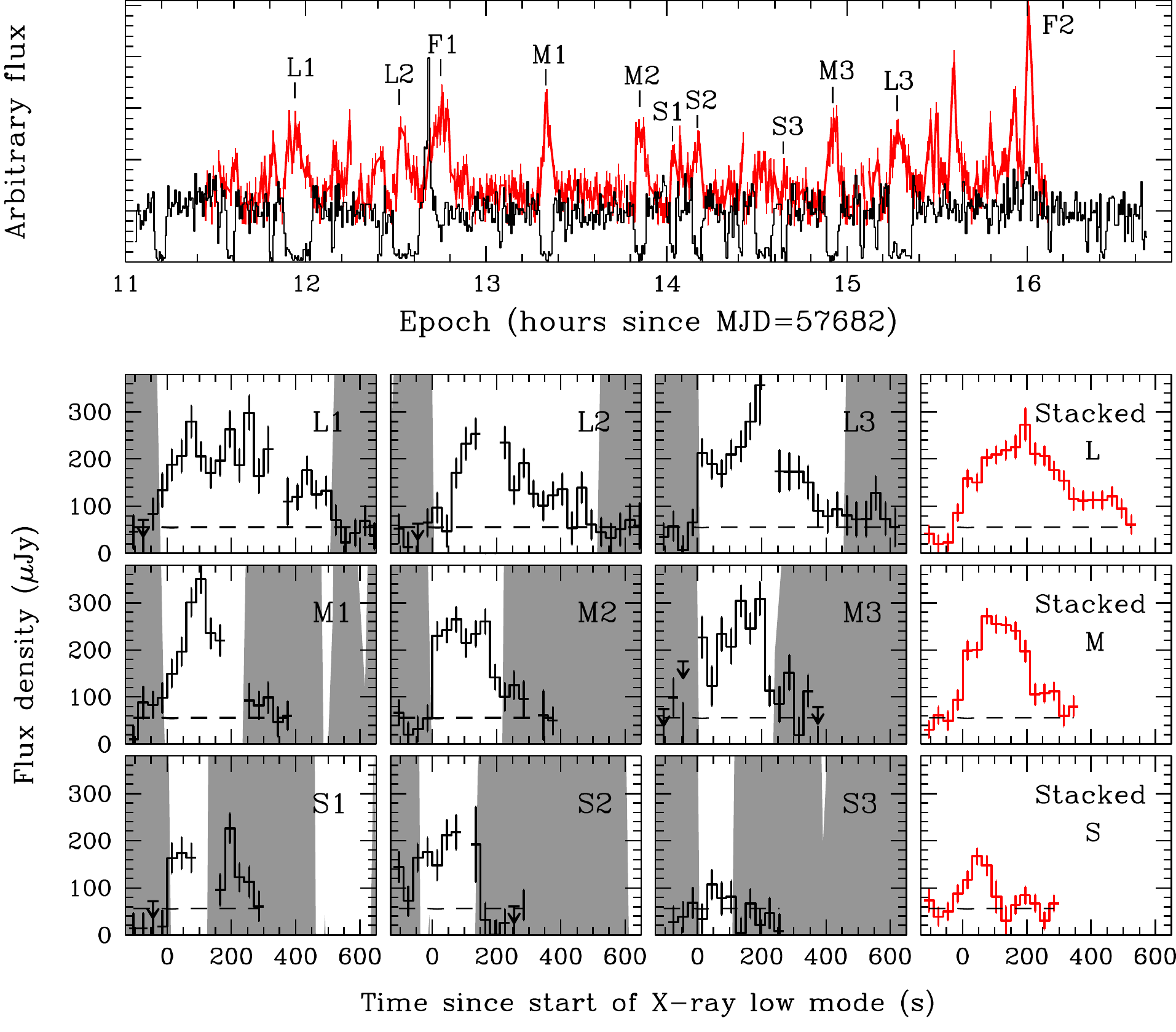}
\caption{\label{fig:long_flares} Close-up radio light curves of the brightness enhancements that coincide with nine X-ray low modes in the 8--12 GHz band, which are labeled in the full light curve (uppermost panel). Gaps in the radio light curve correspond to times when the VLA was observing a calibration source, while upper limits are plotted when the source was not detected above the background.  The grey shaded regions mark the X-ray high mode intervals and the rightmost panel in each row shows a stack of the three low mode light curves of comparable duration.}
\end{figure*}

As noted previously, the one prominent X-ray flare in the \textit{Chandra} time series lasting $\sim$2 minutes at UT 12:40 is followed by a delayed and longer-lasting radio flare (labeled F1 in Figure~\ref{fig:long_flares}).  Since we only have one such flare and cannot stack radio data as we can for the low modes, our sensitivity is too limited to show a significant variation in the spectral index over the course of the flare.  When the radio flare is divided into a rising phase (UT 12:40:48 -- 12:44:48) and decaying phase (UT 12:44:48 -- 12:48:36), the spectral index is $\alpha = 0.4 \pm 0.5$ in the rising phase, and $\alpha = 0.1 \pm 0.5$ in the decaying phase.  While this is consistent with a transition from inverted to steep (as is seen in the X-ray low modes, and would be expected for an expanding region of synchrotron-emitting plasma), it is also consistent with a constant spectral index (as is seen in, e.g., the X-ray high mode).

\subsection{Comparison of X-ray low modes as a function of duration}

\begin{figure}[t!]
\includegraphics[angle=0,width=0.42\textwidth]{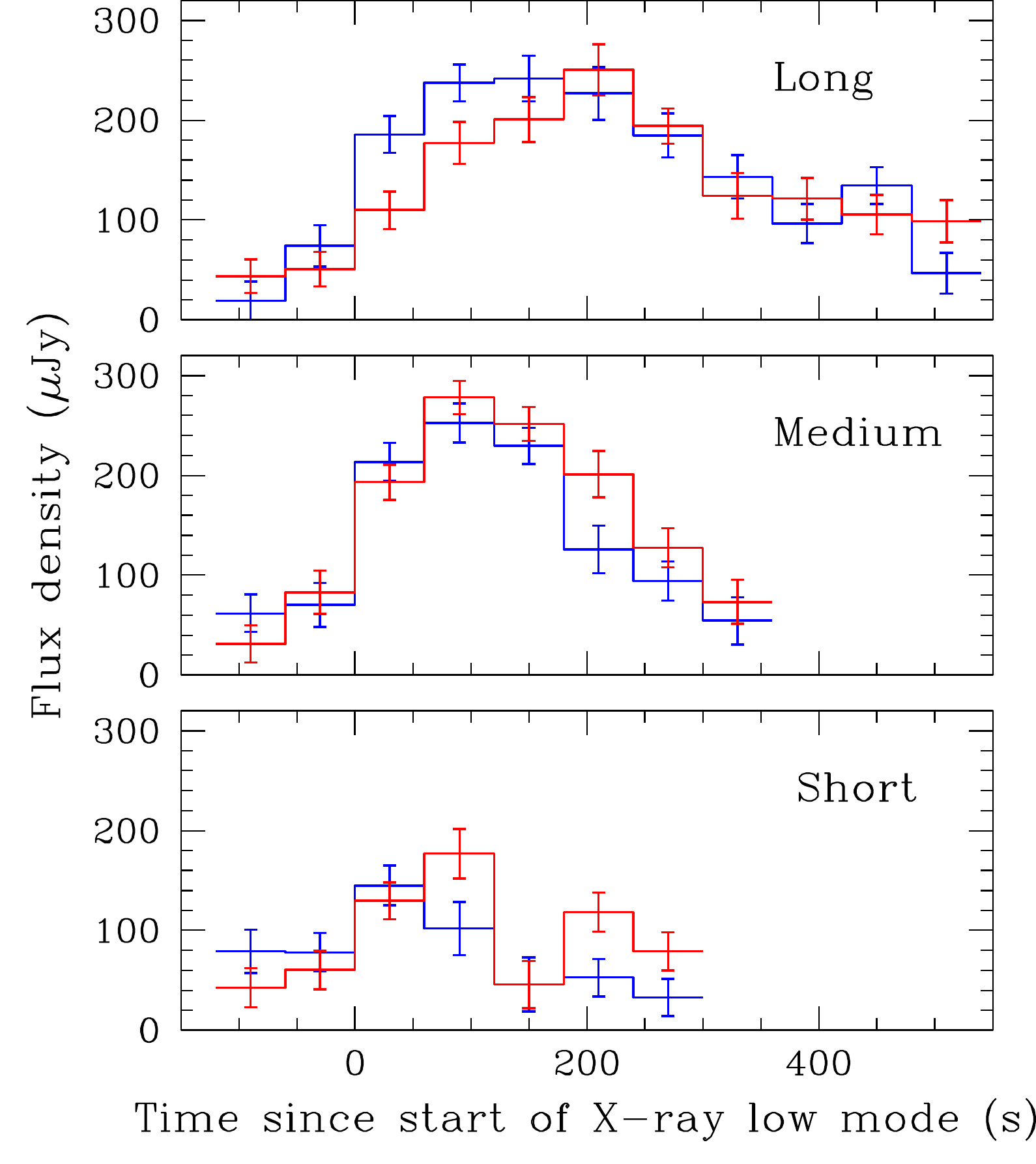}
\caption{\label{fig:lowmodeflares_dualfreq} \textit{Top panel:} Stacked radio light curves of the flares that coincide with the three longest X-ray low modes (labeled L1 through L3 in Figure~\ref{fig:long_flares})  in the 8-10 GHz (red) and 10--12 GHz (blue) bands. \textit{Middle panel:} Stacked radio light curves for the three low modes of duration $\sim$200~s (labeled M1 through M3 in Figure~\ref{fig:long_flares}). \textit{Bottom panel:} Stacked radio light curves of the three shortest X-ray low modes (labeled S1 through S3 in Figure~\ref{fig:long_flares}). For all three cases, the zero time reference corresponds to the start of the X-ray low mode interval.}
\end{figure}

Although the individual low mode radio brightening events do not have sufficient S/N to determine the radio spectral evolution, by combining multiple intervals it is possible to place informative constraints. From the 23 low mode intervals identified in the X-ray data, 18 have strictly simultaneous VLA coverage.  Seven of these have a duration of less than 60 seconds, making them very difficult to resolve with our 30 s radio data, leaving 11 low mode intervals with durations $75-500$ s.  For the purposes of investigating the effects of low mode duration on the radio emission, we divide these into three bins: $88 \pm 12$ s, $200 \pm  15$ s, and $460 \pm 40$ s.  Two low modes, of duration $126$ s and $277$ s, could not be included in any bin without broadening the bin too far, and were discarded from the following analysis.  For each bin (long, medium, and short) we stack the three included individual low modes, using the high $\rightarrow$ low mode transition boundary as a reference point. The results are shown in Figure~\ref{fig:long_flares}, with the L, M, and S labels assigned to the long, medium, and short flares, respectively.  Finally, to highlight the spectral evolution during outbursts we reproduce the stacked images shown in the right column of Figure~\ref{fig:long_flares} with the upper and lower halves of the observed frequency range (8--10 and 10--12 GHz) plotted separately, reducing the temporal resolution to $60$~s to maintain sensitivity (see Figure~\ref{fig:lowmodeflares_dualfreq}). We discuss the findings from the analysis of these light curves in Section 5.

\section{Revisiting the X-ray Binary Radio-X-ray Luminosity Relation}

Black hole (BH) low-mass X-ray binaries (LMXBs) exhibit a well-defined relation between their X-ray and radio luminosities of $L_R\propto L_X^{0.61}$ \citep{2014MNRAS.445..290G}. Neutron stars (NSs) at high luminosities $L_X\gtrsim 10^{36}$ erg s$^{-1}$, on the other hand, may follow two possible correlations, $L_R\propto L_X^{0.7}$ or $L_R\propto L_X^{1.4}$, depending on whether they are in a radiatively efficient or inefficient regime \citep{2006MNRAS.366...79M}. X-ray and radio flux measurements obtained for the three confirmed accreting transitional MSPs, PSR J1023+0038 \citep{2015ApJ...809...13D}, XSS J12270$-$4859 \citep{2011MNRAS.415..235H}, and J1824$-$2452I \citep{2013Natur.501..517P}, have established that X-ray binaries containing MSPs\footnote{The tMSP candidate 1RXS J154439.4$-$112820 \citep[3FGL J1544.6$-$1125;][]{2015ApJ...803L..27B} falls in the same region as the confirmed tMSPs (A.~Jadoand et al.~in preparation).},  can have much higher radio luminosities than predicted from extrapolating the $L_X-L_R$ relation for high accretion rate neutron star LMXBs from a limited range of $L_X$ \citep{2011MNRAS.415.2407M}.  At $L_X\lesssim10^{34}$ ergs s$^{-1}$, the time-averaged radio luminosities of tMSPs are less than an order of magnitude below the BH relation.

\begin{figure*}[t!]
\begin{center}
\includegraphics[angle=0,width=0.75\textwidth]{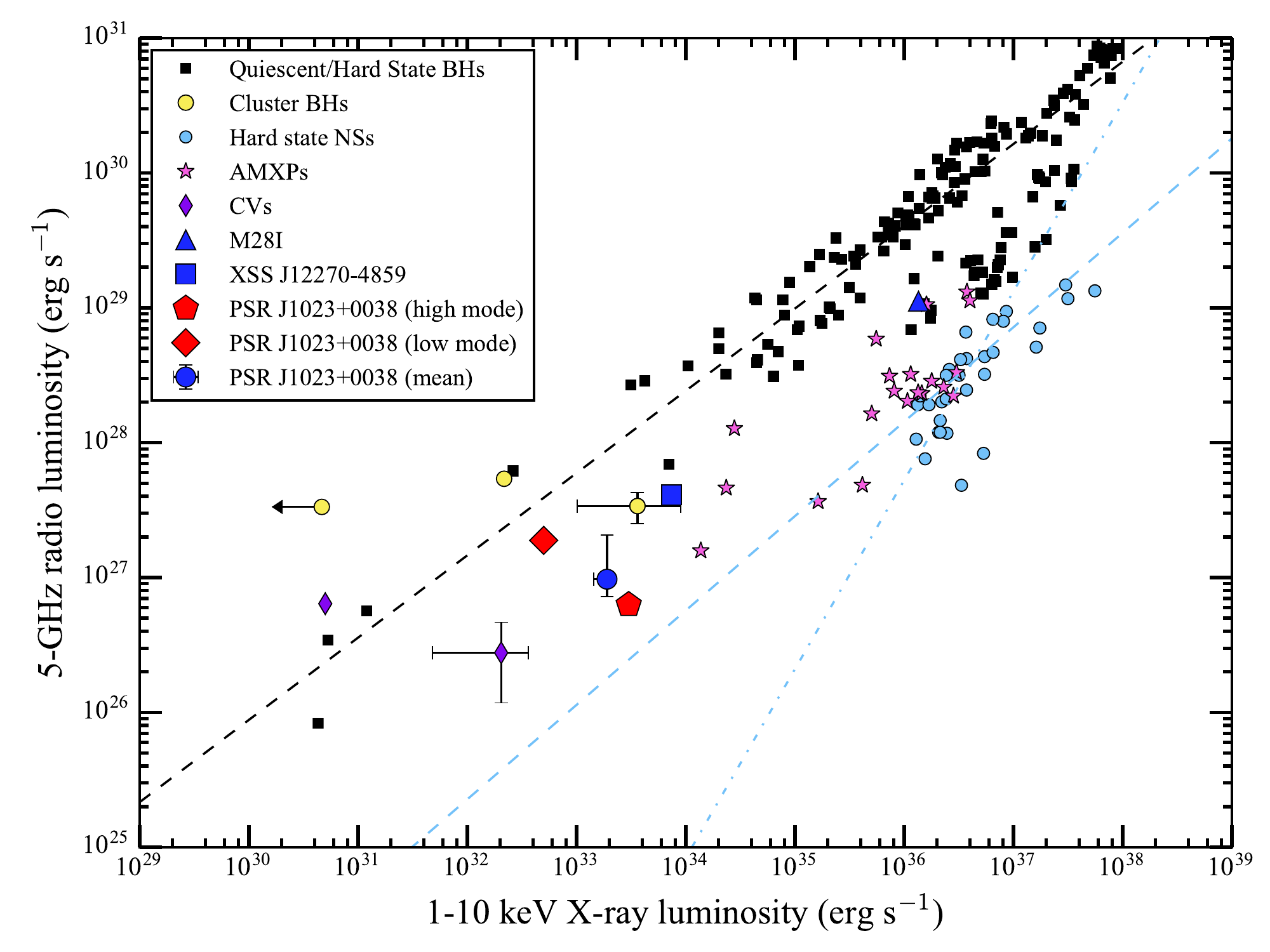}
\caption{\label{fig:lrlx} The radio -- X-ray correlation
for accreting black holes (black points), globular cluster black hole candidates (yellow circles), cataclysmic variables (purple diamonds),  and neutron stars (rest of colored points).
The black hole LMXBs relation $L_R \propto L_X^{0.6}$  is shown by the dashed black line indicative of radiatively inefficient accretion, while neutron star LMXBs are generally less radio bright but their $L_R-L_X$ relation is not well determined.  Transitional pulsars (dark blue) are brighter in the radio for a given X-ray luminosity compared to other varieties of neutron star LMXBs.  The simultaneous
\textit{Chandra}+VLA data points for PSR J1023+0038 (red) show that in the X-ray low mode, this system is indistinguishable from a black hole LMXB. }
\end{center}
\end{figure*}

Figure~\ref{fig:lrlx} shows the  $L_X-L_R$ plane for NS and BH LMXB systems including recent measurements for EXO 1745$-$248 \citep{2016MNRAS.460..345T}, 1RXS~J180408.9$-$342058 \citep{2017MNRAS.470.1871G}, three AMXPs plus Cen X--4 \citep{2017arXiv170505071T}, and the BH LMXB candidates in the globular clusters  M22 \citep{2012Natur.490...71S}, M62 \citep{2013ApJ...777...69C},  and 47 Tuc \citep{2015MNRAS.453.3918M,2017MNRAS.467.2199B}. We also add points for PSR J1023+0038 corresponding to the high and low modes, as well as the mean $L_X$ and $L_R$ values, all computed assuming the parallax distance of $1.368^{+0.042}_{-0.039}$ pc \citep{2012ApJ...756L..25D}.   An important consequence of the anti-correlated behavior of PSR J1023+0038 is that, in the X-ray/radio luminosity plane, the system makes excursions close to the black hole X-ray binary relation. 

While PSR J1023$+$0038 spends only $\sim$20\% of the time in the X-ray low mode, the duty cycle of this mode in other tMSPs can be significantly higher. 
For instance, \textit{Chandra} observations of the tMSP PSR J1824$-$2452I from 2008 showed that this system can reside in a single low mode interval for $\sim$10 hours \citep{2013Natur.501..517P,2014MNRAS.438..251L}. 
This certainly complicates the use of the $L_X-L_R$ relation to identify candidate BHs and invites a close re-examination of globular cluster BH candidates \citep[see][]{2012Natur.490...71S,2013ApJ...777...69C}, where the ratio of X-ray to
radio luminosity obtained from a single short observation was a principal criterion for classifying them as
such.  The use of radio versus X-ray luminosity as a diagnostic for BH LMXBs could be bolstered when large and rapid variability of the type exhibited by tMSPs can be ruled out, and/or once a significant number of observations show consistent results.  The anti-correlated X-ray and radio behavior of  PSR J1023$+$0038  also highlights the necessity for truly simultaneous X-ray and radio observations when attempting to establish the true nature of the accreting compact object.

\section{Discussion}

The simultaneous X-ray/radio data presented here provide valuable new pieces of information on PSR J1023+0038 and, by extension, the tMSP class. The most obvious feature, as shown in Figure~\ref{fig:full}, is the striking anti-correlation of X-ray and radio flux in the high and low modes: the X-ray high modes show X-ray pulsations and low radio luminosity, but when the source switches into its low modes, the X-ray pulsations disappear and the radio emission increases rapidly and dramatically.  Figure~\ref{fig:long_flares} also shows zoomed views of individual low modes, and higher S/N stacks of radio light curves during low modes of similar durations.  Some common signatures can be discerned: generally, the radio rises quickly, begins to decline before the end of the low mode, and returns to quiescence within $30-60$\,s of the transition back to the X-ray high mode.  Shorter low modes appear to rise faster than longer low modes and can reach comparable peak luminosities, although the shortest flares have consistently fainter peaks.  The radio luminosity begins to decline back to the high-mode level just before the transition back to the high mode, although in the longest low modes, the radio has already declined to close to the quiescent value. In the shorter low modes on the other hand, the radio emission is still bright and the `cut-off' at the end of the low mode is more apparent.

During the X-ray low modes, there is good evidence that the radio spectrum evolves from initially inverted to steep (see Table 1 and Figure~\ref{fig:lowmodeflares_dualfreq}) -- reminiscent of a classical evolving synchrotron spectrum, going from self-absorbed to optically thin on relatively short timescales.  This occurs on timescales of several minutes. In contrast, when in the X-ray high mode, the radio emission is normally steady at $\sim$55 $\mu$Jy.  However, several instances can be seen in which a radio flare appears to occur during an X-ray high mode with no discernible or only weak enhancement in X-ray emission (see F2 in Figure~\ref{fig:long_flares}). Finally, the bright X-ray flare event appears to be followed by a radio flare (F1 in Figure~\ref{fig:long_flares}) that exhibits classic evolving synchrotron features of accretion-driven outflows: a time delay relative to the X-ray event and spectral evolution consistent with expanding plasma (although, as noted, the low signal--to--noise ratio of the radio data also permits a non-evolving spectrum.) 

The low radio luminosity during the X-ray high mode, when active accretion onto the star occurs as evidenced by the coherent X-ray pulsations, can be plausibly interpreted as a weak persistent compact self-absorbed synchrotron jet, commonly observed in both NS \citep{2006ApJ...643L..41M,2010ApJ...710..117M} and BH \citep{2000MNRAS.312..853F,2001MNRAS.327.1273S,2005MNRAS.356.1017G}  LMXBs. Ejection of a denser discrete plasma cloud can explain the intense X-ray flare and the associated delayed radio flare, which is an expected feature of intermittent jet production \citep[see, e.g.,][]{2003ApJ...597.1023V,2004MNRAS.355.1105F}. On the other hand, a similar jet launching scenario seems improbable for the rapid low mode radio brightness increases. In particular, in the conventional accretion/jet production paradigm in X-ray binaries \citep{1982MNRAS.199..883B}, the enhancement in accretion rate, observed as increased X-ray flux, is associated with a commensurate increase in jet power, manifested as brighter radio emission in the hard spectral state. However, the anti-correlated X-ray/radio behavior seen in PSR J1023$+$0038 at X-ray luminosities characteristic of a hard or quiescent state implies that a decline in accretion rate is associated with an enhancement instead of a reduction in plasma outflows. Furthermore, a scenario involving radio emission from accretion-driven outflow at large distances from the compact object cannot be reconciled with the fact that the initial rise and decline of the radio flares during the low modes always closely matches the X-ray low mode interval. In particular, once ejected, material in a jet-like outflow no longer maintains causal contact with events occurring near the pulsar, and thus would have no information regarding the X-ray low$\rightarrow$high mode transition, which is caused by events within the pulsar magnetosphere -- inner disk interface. This leads us to conclude that if these radio flares are due to expanding/outflowing synchrotron-emitting plasma (as the spectral evolution suggests), the bulk of this emission must arise from plasma in the immediate vicinity of the inner disk region.    

The cessation of coherent X-ray pulsations at the high$\rightarrow$low X-ray mode transition\footnote{As reported in \citet{2015ApJ...807...62A}, the pulsed fraction limit in the X-ray low mode ($2.4\%$ at 95\% confidence) is well below the observed pulsed fraction in the high mode ($8.1\%$).} further implies that the event that leads to the production of the radio flare is also (directly or indirectly) responsible for temporarily disrupting the accretion flow onto the pulsar. \citet{2016A&A...594A..31C} have applied a multi-component model, including a radiatively-inefficient disk, to the extensive archival \textit{XMM-Newton} X-ray data of PSR~J1023$+$0038. Based on this, they deduce an inner disk radius of $\sim$21 km in the X-ray high mode and $\sim$200 km in the X-ray low mode. This finding is consistent with an interpretation in which the accretion flow is pushed out beyond the light cylinder, perhaps by ejection/expansion of plasma originating in the pulsar magnetosphere, causing a high$\rightarrow$low X-ray mode transition.

\citet{2015ApJ...807...62A} and \citet{2015ApJ...807...33P} proposed a basic explanation for the peculiar X-ray mode switching: in the high modes, magnetically-channeled accretion onto the neutron star surface occurs, and when the source switches to a low mode, propeller-mode accretion takes over. This model can account for the anti-correlation we see: in the low modes there is propeller-mode accretion and material is ejected, producing a radio-emitting jet, while in the high modes the material reaches the surface instead, producing X-ray pulsations. Of course, this model is oversimplified and has problems on both theoretical and observational levels. From a purely theoretical standpoint, between accretion and ejection we expect a ``weak propeller'' regime \citep{2006ApJ...646..304U}, where the centrifugal barrier prevents accretion but the magnetically-coupled material at the inner edge of the disc does not have escape velocity and cannot be ejected. Further, the X-ray luminosity even in the high modes is far below that predicted to be necessary for matter to reach the co-rotation radius; this can possibly be explained in terms of ``trapped disk'' scenarios \citep{2012MNRAS.420..416D} but it is unclear how ejection fits into these pictures. Moreover, MHD simulations of accretion in this regime show rapid switching between two modes but with a \emph{positive} correlation: short episodes of accretion are accompanied by ejection, and these are interspersed with episodes where neither occurs; these episodes also occur on drastically shorter timescales (of order a few rotation cycles of the star) than we observe. Finally, the rapid disappearance of the radio emission we report here poses challenges for jet-type models: how does the jet disappear so abruptly? Clearly more nuanced explanations are necessary.

Alternative interpretations for the multi-wavelength phenomenology of PSR J1023+0038 and the other accreting tMSPs have centered on the assumption of an active rotation-powered pulsar wind \citep[see, e.g.,][]{2014ApJ...785..131T,2014MNRAS.438..251L,2014MNRAS.444.1783C,2016A&A...594A..31C}. The spin-down rate of PSR J1023$+$0038 has been measured to be $26.8\pm0.4$\% higher in the accretion state as compared to the radio pulsar state \citep{2013arXiv1311.5161A,2016ApJ...830..122J}. This is an indication that the pulsar spin-down still proceeds largely unaffected by the accretion inflow, meaning that even if the pulsar radio emission mechanism is disrupted in the accretion state the bulk of the rotational energy loss is still in the form of a relativistic particle wind and Poynting flux. 

Much of our insight into accreting neutron star systems and MSPs in particular has been gained from observations of high-luminosity, that is, high accretion rate, states. In this regime, the power output of a rotation-powered pulsar relative to accretion power is vastly subdominant and its role in the observed properties is virtually negligible. In contrast, tMSPs reside in a regime where accretion and rotation power are comparable. In the case of PSR J1023$+$0038 in its present low-luminosity accreting state, the spin-down luminosity of $\dot{E}\approx4\times10^{34}$ ergs s$^{-1}$ is only about an order of magnitude greater than the time averaged X-ray luminosity, which is presumably powered mostly by accretion. Therefore, for an accretion efficiency of $GM/c^2R\sim0.1$, rotation and accretion power are comparable.   
In light of this, it is perhaps not surprising that the classical accretion inflow/jet outflow model cannot offer a satisfactory, self-consistent explanation of the observed X-ray/radio behavior of PSR J1023+0038 due to the lack of a key ingredient, namely, an active, rapidly spinning rotation-powered pulsar that acts as an additional reservoir of energy for driving outflows. Recently, \citet{2016ApJ...822...33P}  and \citet{2017MNRAS.469.3656P}  modeled the interaction of an active magnetosphere with a matter-dominated disk with a fixed velocity field, and finite thickness and conductivity. The simulations do indicate that for accreting MSPs the pulsar wind outflow may power the radio jets. However, they also predict a dramatic increase in spin-down of the neutron star due to opening of additional magnetic field lines, which is not observed in PSR J1023+0038 \citep{2016ApJ...830..122J}, implying that in reality the magnetic field is much less efficient at coupling the star and disk. 
\citet{2017arXiv170806362P} furthered these simulations to  interaction of a force-free magnetosphere with a full-MHD accretion disk in a general relativistic regime for the first time. The set of simulations produces four magnetic field strength dependent regimes, including two that may account for the X-ray modes in tMSPs: (1) a regime in which the pulsar wind inhibits the accretion flow from moving past the light cylinder but drives shocks through it giving rise to X-ray emission, and (2) a regime in which the accreting matter is able to penetrate the light cylinder but is occasionally ejected by the rotating magnetic field. However, these simulations only span a real-time duration of $\sim 0.35$\,s (assuming a NS mass of $1.4$\,M$_\odot$), in contrast to the observed X-ray mode-switching timescales of $\sim 10$\,s. Although these simulations do not yet predict the detailed multi-wavelength behavior seen in PSR J1023+0038, they nevertheless represent important advances towards an understanding of accretion onto rotation-powered MSPs.

Under the assumption of an active pulsar, one possible interpretation of the anti-correlated X-ray/radio variability involves the rapid discharge of plasma from the pulsar magnetosphere or the inflation of a short-lived compact pulsar wind nebula. In this scenario, due to circumstances that remain to be understood, outflows from the pulsar are occasionally able to expand past the pulsar light cylinder in the orbital/accretion disk plane and likely out to much larger distances above and below the disk. Presumably, accretion commences again once the rapidly expanding plasma cools sufficiently. The duration of an X-ray low mode/radio brightening interval probably depends on the initial conditions of the plasma ejection event.
 
If outflows from the pulsar are responsible for the observed X-ray mode switching and associated radio brightness enhancements a crucial question that arises is: What precipitates these episodic eruptions?
The highly reproducible X-ray mode-switching behavior is not seen in other varieties of accreting systems but is reminiscent of certain phenomenology associated with rotation-powered pulsars, such as radio mode switching and nulling \citep{1970Natur.228...42B,2010MNRAS.408L..41T,2013Sci...339..436H,2014A&A...572A..52B,2016ApJ...831...21M}. While such phenomena likely occur in the absence of external drivers such as accreting material, they demonstrate that abrupt reconfiguration of the pulsar magnetosphere between discrete metastable states is possible.  Although the disk-magnetosphere interaction is poorly understood, the relative energy budgets of accretion and spin-down as well as the phenomenology suggest that pulsar electrodynamics may govern the physical processes responsible for the peculiar switching behavior. The pulsar may also be responsible for the extraordinary stability of the high mode X-ray flux, which implies a near constant accretion rate onto the star over many years \citep{2015ApJ...806..148B,2016ApJ...830..122J}, perhaps through an as-yet-unknown flow regulation mechanism imposed by the pulsar magnetosphere.

\section{Conclusions}
We have presented coordinated X-ray and radio observations of PSR J1023+0038 in its low-luminosity accreting state. For the first time, we conclusively establish the presence of strongly anti-correlated variability in the two bands. In particular, the radio flux density shows a rapid rise that commences at roughly the same time as the drop from a high to a low X-ray flux mode, peaking at most minutes later before beginning a decline back towards the ``quiescent" level seen during the X-ray high mode.  However, if this decline is not completed by the low$\rightarrow$high mode transition, a relatively abrupt drop back to the quiescent level is seen at around this time.  This phenomenology implies a causal connection that is difficult to explain by any standard inflow/outflow models of accretion onto a compact object.  The radio spectral evolution during a low mode (inverted early in the low mode, steep late in the low mode) is particularly difficult to explain in light of this causal connection, since it would be most easily interpreted as an expanding synchrotron-emitting plasma, whose causal connection to the accretion flow (as probed by the X-ray) is difficult to fathom.

We argue that the active rotation-powered pulsar plays a key role in the production of the anti-correlated X-ray/radio variability. The radio brightness enhancements accompanying X-ray low modes might arise from plasma outflows driven by the active rotation-powered pulsar instead of the accretion inflow, while the mode switching is the result of bistable reconfigurations of the pulsar magnetosphere, perhaps induced by the action of the accretion flow.

The simultaneous X-ray and radio data further reveal that, during the X-ray low modes, PSR J1023$+$0038 has a position in the $L_X-L_R$ diagram that makes it indistinguishable from a BH LMXB. This finding opens the possibility that some candidate low-luminosity BH LMXBs observed only for a short period of time may actually be tMSP impostors caught in an X-ray low mode, which implies that additional lines of observational evidence are needed to unambiguously establish the nature of the compact object \citep[see, e.g.,][for an illustrative example]{2017MNRAS.467.2199B}.

\acknowledgments
SB was funded in part by NASA Chandra Cycle 17 Guest Observer Program grant GO6-17034X awarded through Columbia University and issued by the Chandra X-ray Observatory Center, which is operated by the Smithsonian Astrophysical Observatory for and on behalf of NASA under contract NAS8-03060.  ATD and JCAMJ are the recipients of Australian Research Council Future Fellowships (FT150100415 and FT140101082).  AMA is an NWO Veni fellow.  JWTH, AP, and CD acknowledge support from NWO Vidi fellowships. JWTH and AJ acknowledge support from the European Research Council under the European Union's Seventh Framework Programme (FP/2007-2013) / ERC Grant Agreement nr. 337062.  The National Radio Astronomy Observatory is a facility of the National Science Foundation operated under cooperative agreement by Associated Universities, Inc. This research has made use of the NASA Astrophysics Data System (ADS), the arXiv, and software provided by the Chandra X-ray Center (CXC) in the application package
CIAO.

\facilities{CXO, VLA, XMM} 
\software{CIAO, CASA, SAS}

\bibliographystyle{yahapj}
\bibliography{J1023_refs}


\end{document}